# Pre-notched dog bone small punch specimens for the estimation of fracture properties

I.I. Cuesta [1], A. Willig [2], A. Díaz [1], E. Martínez-Pañeda [3], J.M. Alegre [1]

[1] Structural Integrity Group, Universidad de Burgos, Escuela Politécnica Superior, Avda. Cantabria s/n, 09006 Burgos. SPAIN

[2] AESA, Barreiro 2871, Ruta Provincial 52, Km 32,5, B1804EYA Canning, Ezeiza, Buenos Aires. ARGENTINA

[3] University of Cambridge, Department of Engineering, Trumpington Street, Cambridge CB2 1PZ, UNITED KINGDOM

e-mail: iicuesta@ubu.es

**ABSTRACT**

In recent years, the pre-notched or pre-cracked small punch test (P-SPT) has been successfully used to estimate the fracture properties of metallic materials for cases in which there is not sufficient material to identify these properties from standard tests, such as CT or SENB specimens. The P-SPT basically consists of deforming a pre-notched miniature specimen, whose edges are firmly gripped by a die, using a high strength punch. The novelty of this paper lies in the estimation of fracture properties using dog-bone-shaped specimens with different confinement levels. With these specimens, three confinement variations have been studied. The results obtained enable the establishment of a variation of fracture properties depending on the level of confinement of each miniature specimen and selection of the most appropriate confinement for this goal.

## 1. INTRODUCTION

In recent decades, diverse alternative tests have been used for the mechanical characterization of materials when standard tests cannot be carried out for various reasons. For cases in which there is a shortage of material, the small punch test (SPT) has been established as a viable alternative for mechanical characterization of a material. The test basically consists of deforming a miniature specimen ($10x10x0.5\,mm$), whose edges are firmly gripped by a die, using a high strength punch. During the test, the values of applied load and displacement of the punch are noted, and after appropriate treatment of the test data, a load-displacement curve of the punch is obtained. This test was developed in the nuclear industry in the eighties [1] and has since been successfully used on numerous occasions when there is not sufficient material to carry out standard tests to obtain the mechanical properties of materials, such as in the case of welds or irradiated material [2-5]. The SPT has also been used to estimate material fracture properties [6,7], and in recent years, pre-cracked [8] or pre-notched [9,10] specimens have gained popularity. Some authors have used FEM ductile damage models to simulate specimen behaviour up to failure [11-13]. This test has also been successfully used in other types of studies, such as corrosion [14,15] or creep studies [16,17]. The experimental setup can be designed according to the CEN code of practice for small punch testing [18].

To date, all SPT investigations have employed specimens that are completely embedded in the die of the tooling; that is, the confinement level of the specimen is high. However, what happens if the level of confinement varies? The novelty of this paper is the analysis of the effect of the degree of confinement of a specimen on its fracture behaviour. The confinement is varied by modifying the geometry of the specimen from a conventional square shape to a dog-bone shape, while maintaining the same thickness ($0.5\,mm$). With these specimens, three confinement variations have been studied. The obtained results enable the establishment of a variation of fracture properties depending on the level of confinement of each miniature specimen and a selection of the most suitable confinement for this goal.

## 2. MATERIALS

The material used in the present study is a series 5000 aluminium-magnesium alloy, which was annealed to assess its lowest strength and then hardened by deformation to assess its final mechanical properties. The material is denoted AW 5754-H111, and a sheet of material with a thickness of $0.512\,mm$ and $R_z$ roughness of $2.386\,\mu m$ was utilized. Therefore, no additional polishing of the miniature punch specimens was necessary as the roughness levels are below the 2.5 $\mu$m threshold established by the CEN code of practice for small punch testing [18].

Due to its conformability and ease in welding and laser cutting, this alloy is commonly used in manufacturing containers for liquids, pressure vessels, tanks for transporting hot loads and pipes for heat exchangers and naval applications, among others. Its mechanical strength is medium/high, with excellent resistance to corrosion. However, this material tends to suffer intercrystalline and corrosion cracking under stress after inadequate heat treatment (welding). The chemical composition of this type of alloy is defined in Table 1.

### 3. METHODOLOGY

To study the effect of varying the degree of confinement of P-SPT specimens on their fracture behaviour, pre-notch dog-bone-shaped SPT specimens were machined and characterized by the specimen width $W$ (Figure 1); all specimens were cut and pre-notched using a laser.

The P-SPT was carried out at room temperature with a punch diameter of $d_p = 2.5\,mm$, and the hole in the lower die had a diameter of $D_d = 4\,mm$ with a fillet radius of $r = 0.5\,mm$ (Figure 2). The punch drop rate was $v = 0.5\,mm/min$.

To obtain the fracture properties from P-SPT, a multi-specimen approach was used based on the method developed by Landes and Begley [19], who were the first to experimentally measure the *J-integral*. This method enables to compute an energy quantity analogous to the J-integral for P-SPT. This is achieved by testing a series of specimens of the same size, geometry and material but with different crack lengths. Each specimen is deformed, and a load-displacement curve is obtained. The area under a given curve is equal to the energy absorbed by the specimen ($U$). When a crack on the specimen initiates and extends from a value of $a_1$ to $a_2 = a_1 + da$, the *J-integral* can be obtained as

$$J = -\frac{1}{t}\left(\frac{\partial U}{\partial a}\right)_\Delta = -\frac{1}{t}\left(\frac{U_2 - U_1}{a_2 - a_1}\right) \quad (1)$$

where $U_1$ and $U_2$ are the energy under the load-displacement curve of two different specimens with crack lengths $a_1$ and $a_2$, respectively, and $t$ is the thickness of the specimen.

If the $J$ value is evaluated up to fracture, a *J-integral* initiation value can be obtained. Because crack initiation in P-SPT specimens takes place at the maximum load, the energy $U_1 - U_2$ under the load displacement evaluated until $P_{max}$ represents the energy used to extend the crack from $a_1$ to $a_2$. A graph of this approach is shown in Figure 3.

It should be noted that P-SPT specimens do not have the same behaviour as CT, SENB or DDEN-T standard specimens when the (uncracked) ligament length ($L$) approaches zero. For example, in a standard specimen, if $L$=0, the load also goes towards 0 as the specimen is separated into two halves.

However, in a P-SPT specimen, the load is non-zero when *L*=0, since the punch keeps plastically deforming the two halves of the specimen to pass through them.

In other words, it is necessary to subtract the fixed value of the energy corresponding to the P-SPT specimen with *L*=0 from each specimen's energy value to ensure that P-SPT specimens resemble standard specimen behaviour when *L*=0. Here, this aspect is already taken into account when applying Equation 1 since the result of $U_1 - U_2$ corresponds to the energy variation between 1 and 2.

## 4. RESULTS AND DISCUSSION

Figure 4 shows the most characteristic load-displacement curves obtained via P-SPT for pre-notched dog-bone-shaped specimens. A corresponding SPT curve showing the behaviour of a conventional specimen ($10x10\,mm$) has also been included. It is clear that the different confinement levels affect the SPT curves.

To apply the method presented in the previous section, five miniature specimens have been tested for each specimen width analysed (2, 3 and $4\,mm$), with similar notch values, thus allowing the use of Equation 1. In addition, the exact moment of crack propagation initiation in the P-SPT must be known. From an engineering standpoint and to simplify internal energy calculations, the maximum load point can be used as the starting point. However, previous studies indicate that the initial crack propagation point is reached before the maximum load is applied [8, 20]. Therefore, both scenarios will be analysed.

To identify the appropriate point, the elastic load method [20], in which the slopes of elastic loads are analysed for pre-cracked specimens, was used. The procedure is similar to the conventional P-SPT, with the inclusion of elastic loads in equidistant time periods. The slope recorded in each load represents the flexibility of the specimen, and therefore, its value should change once the crack starts to propagate. An example of a load-displacement curve obtained from this method, which shows the change in the load slope during the P-SPT, is included in Figure 5, corresponding to a miniature dog-bone specimen with a width of $4\,mm$. The results clearly show that in the initial stretch of the curve, the slopes tend to behave linearly as a function of displacement up to 70% of the maximum load, at which point the slopes no longer show linearity. Therefore, crack propagation can be considered to start at this point. To validate this statement, tests interrupted at approximately 70% of the maximum load were performed on different types of specimens. All tests showed that crack propagation effectively starts at approximately 70% of the maximum load.

Therefore, this point ($70\% P_{max}$) and the maximum load point were used to carry out internal energy calculations of miniature pre-notched specimens. For each miniature specimen tested, the energy at both 70% of the maximum load and at the maximum load were calculated based on the area under the curve.

For each group of specimens with the same width, the energy values were compared by applying Equation 1 to all possible pairs of the different specimens. Figure 6 shows the values of fracture toughness as a function of specimen width. Figure 6 also shows the values of fracture toughness assessed for compact specimens of a similar 5000 series alloy [21]. Clearly, the results at 70% of the maximum load best approximate those obtained when using compact specimens. The dispersion is also noticeably smaller than that at maximum load values.

Specimens with a width of $4 mm$ best approximate CT specimens with an estimated fracture toughness of $115 \pm 30 kJ/m^2$ and the compact specimen with the lowest strain triaxiality (CT10), $118 kJ/m^2$. Therefore, the results validate the use of miniature dog-bone specimens to estimate the fracture toughness of a material when there is not sufficient material to perform conventional tests and when the tensile state is close to that of plane stress conditions.

## 5. CONCLUSIONS

In summary, the applicability of miniature dog-bone specimens has been analysed. The results show that miniature specimens with a width of $4 mm$ best estimate fracture toughness results when using an estimation of the energy released in the fracture process.

In addition, the use of the J-integral showed that crack propagation initiation can be assumed to occur at approximately 70% of the maximum load of the specimen and that this assumption reduces the dispersion of the estimated toughness results.

These novel findings for miniature dog-bone specimens further support the use of the SPT as an alternative to conventional tests to estimate the fracture properties of ductile materials under plain stress conditions.

## 6. ACKNOWLEDGEMENTS

The authors gratefully acknowledge financial support from the Ministry of Economy and Competitiveness of Spain through grant MAT2014-58738-C3.

**Table and figure list:**

| %Si | %Fe | %Cu | %Mn | %Mg | %Zn | %V | %Others | %Al |
|---|---|---|---|---|---|---|---|---|
| 0.08 | 0.1 | 0.2 | 0.15-0.45 | 0.8-1.2 | 0.05 | 0.05 | 0.1 | Rest |

**Table 1.** *Chemical composition of AW 5754-H111.*

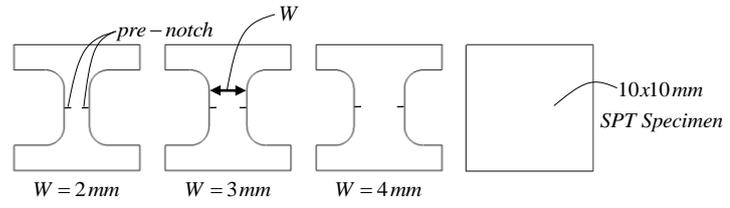

**Figure 1.** *Pre-notch dog bone shaped SPT specimens.*

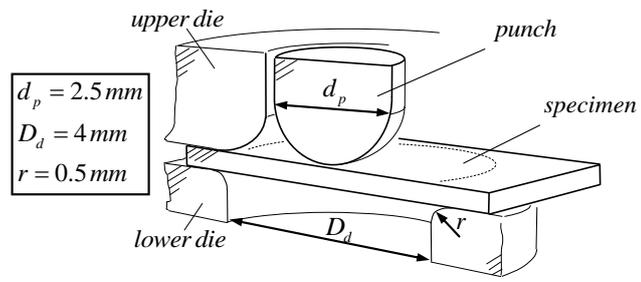

**Figure 2.** *Typical SPT experimental setup.*

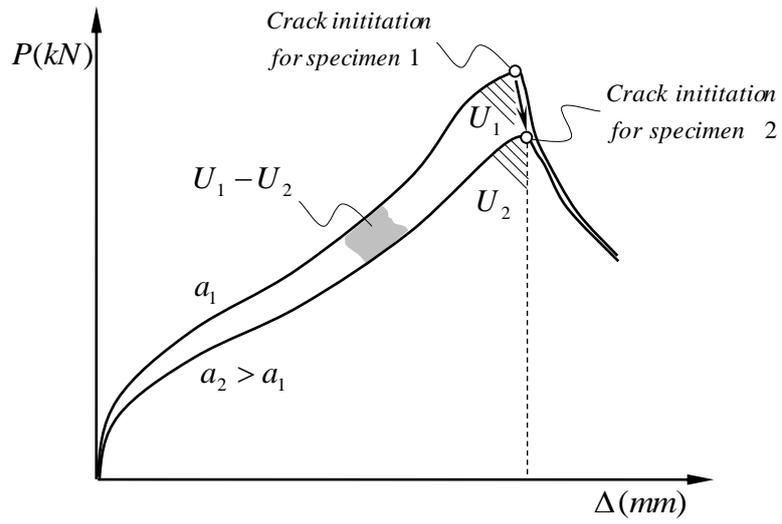

**Figure 3.** *Multi-specimen method for the determination of the fracture energy.*

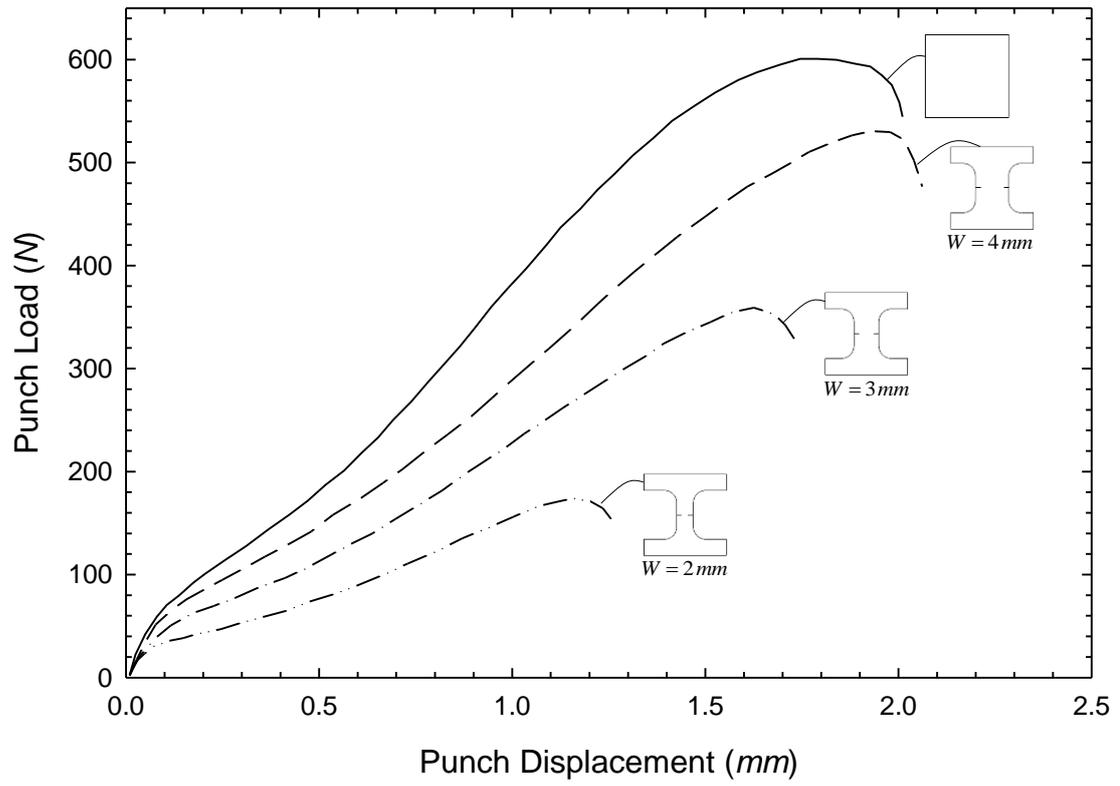

**Figure 4.** *SPT load-displacement curves for different levels of confinement.*

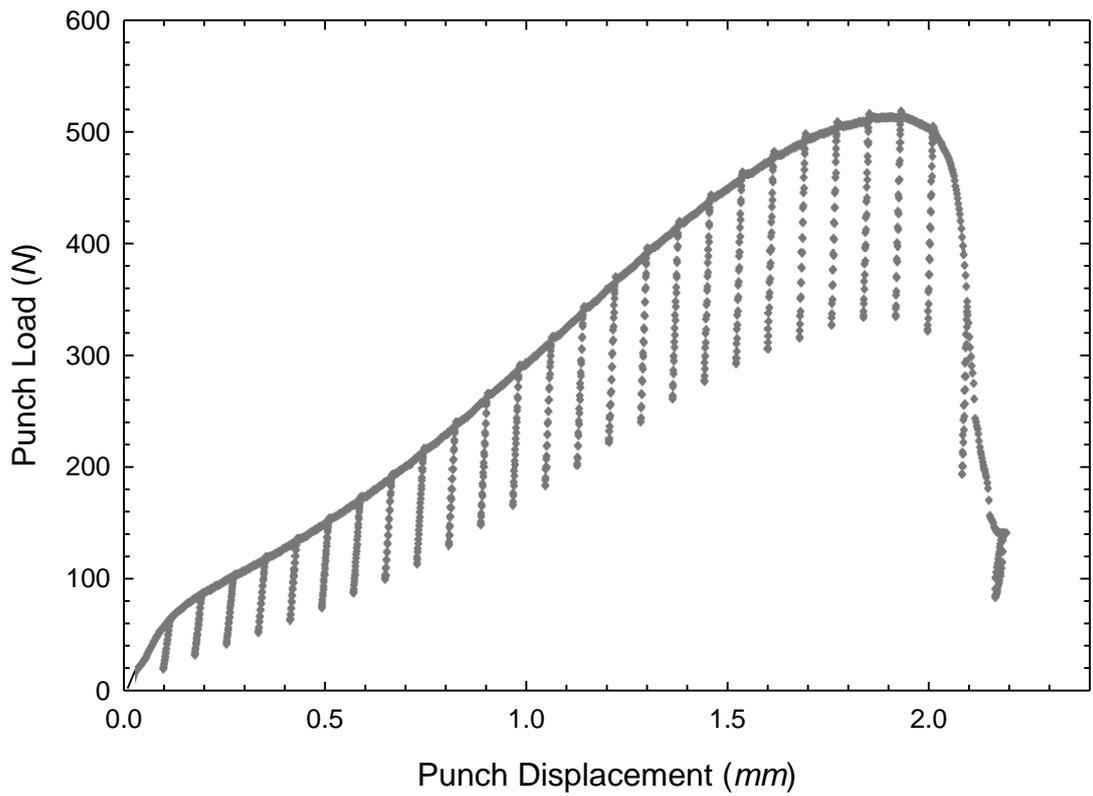
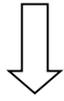
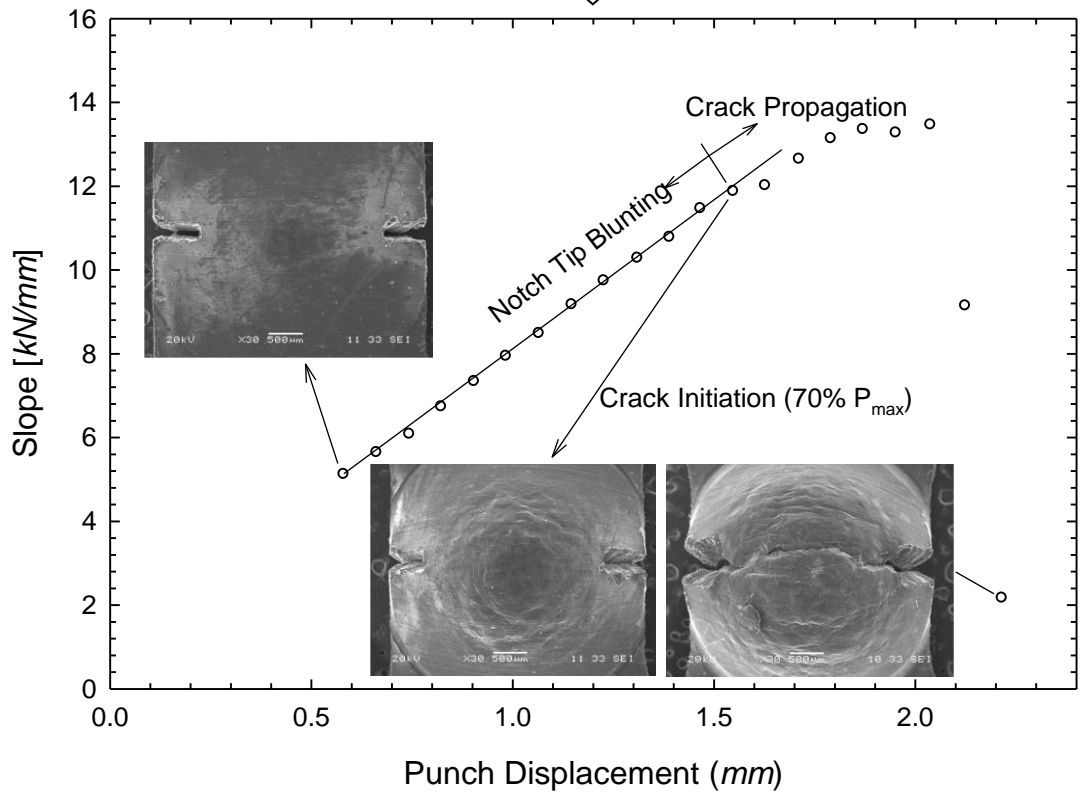

**Figure 5.** *The elastic load method applied to a miniature dog-bone specimen.*

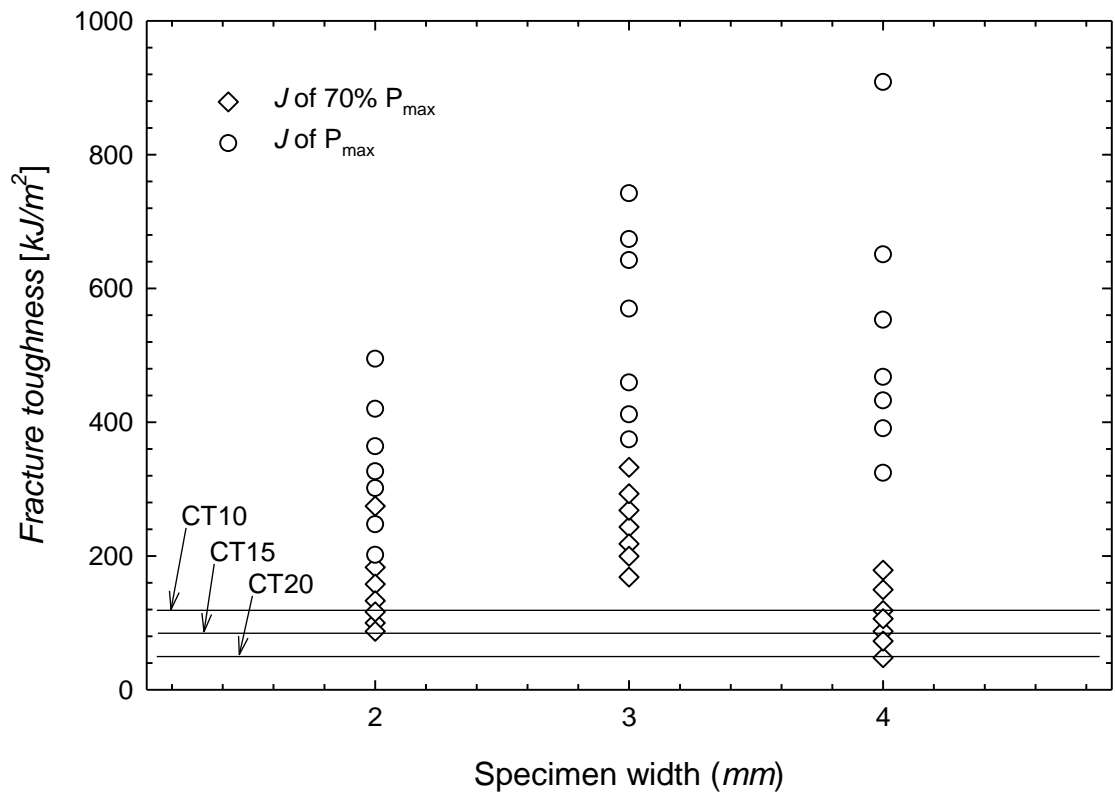

**Figure 6.** *Relationship between fracture toughness and specimen width.*